\documentclass[11pt]{article}\usepackage[]{graphicx}\usepackage[]{color}
\makeatletter
\def\maxwidth{ %
  \ifdim\Gin@nat@width>\linewidth
    \linewidth
  \else
    \Gin@nat@width
  \fi
}
\makeatother

\definecolor{fgcolor}{rgb}{0.345, 0.345, 0.345}
\newcommand{\hlnum}[1]{\textcolor[rgb]{0.686,0.059,0.569}{#1}}%
\newcommand{\hlstr}[1]{\textcolor[rgb]{0.192,0.494,0.8}{#1}}%
\newcommand{\hlopt}[1]{\textcolor[rgb]{0,0,0}{#1}}%
\newcommand{\hlstd}[1]{\textcolor[rgb]{0.345,0.345,0.345}{#1}}%
\newcommand{\hlkwb}[1]{\textcolor[rgb]{0.69,0.353,0.396}{#1}}%
\newcommand{\hlkwc}[1]{\textcolor[rgb]{0.333,0.667,0.333}{#1}}%
\newcommand{\hlkwd}[1]{\textcolor[rgb]{0.737,0.353,0.396}{\textbf{#1}}}%

\usepackage{framed}
\makeatletter
\newenvironment{kframe}{%
 \def\at@end@of@kframe{}%
 \ifinner\ifhmode%
  \def\at@end@of@kframe{\end{minipage}}%
  \begin{minipage}{\columnwidth}%
 \fi\fi%
 \def\FrameCommand##1{\hskip\@totalleftmargin \hskip-\fboxsep
 \colorbox{shadecolor}{##1}\hskip-\fboxsep
     \hskip-\linewidth \hskip-\@totalleftmargin \hskip\columnwidth}%
 \MakeFramed {\advance\hsize-\width
   \@totalleftmargin\z@ \linewidth\hsize
   \@setminipage}}%
 {\par\unskip\endMakeFramed%
 \at@end@of@kframe}
\makeatother

\definecolor{shadecolor}{rgb}{.97, .97, .97}
\definecolor{messagecolor}{rgb}{0, 0, 0}
\definecolor{warningcolor}{rgb}{1, 0, 1}
\definecolor{errorcolor}{rgb}{1, 0, 0}
\newenvironment{knitrout}{}{} 

\usepackage{alltt}

\usepackage{alltt}
\usepackage{amsmath,amscd,amsfonts, thumbpdf}
\usepackage{hyperref}
\usepackage{authblk}

\newcommand{\pkg}[1]{{\normalfont\fontseries{b}\selectfont #1}}
\let\proglang=\textsf
\let\code=\texttt

\title{\pkg{distdichoR}: a \proglang{R} Package  for the distributional dichotomisation of continuous outcomes  }
\author{Odile Sauzet, Universit\"at Bielefeld \and Jannik Rehse, Universit\"at Bielefeld \and Janne Helene Breiding, Universit\"at Bielefeld}
\date{}
\IfFileExists{upquote.sty}{\usepackage{upquote}}{}
\begin{document}

\maketitle

\noindent\rule{\textwidth}{1pt}
\begin{abstract}
We introduce the functions included in the \proglang{R} Packet \code{distdichoR} for the implementation of the distributional method for the dichotomisation of continuous outcomes. While recalling the principle of the method for the distributions for which the method has already been developed, we add a new distribution - the gamma distribution - to the range of distribution available. \\
\textit{Keywords}: dichotomisation, distributional method, cut-point
\end{abstract}
\noindent\rule{\textwidth}{1pt}

\section{Introduction}

The dichotomisation of continuous outcomes, in particular in the health sciences, is commonly done because it makes study results easier to interpret in terms of population at risk or patients needing treatment. However, there are  numerous arguments against it (\cite{Ragland.1992, Royston.2006}), the most obvious one being the lost of most statistical information contained in the data. The distributional method for the dichotomisation of continuous outcomes has been developed to allow comparisons of proportions to complement a comparison of means with equal precision using the underlying parametric distribution of the outcome. The original work was developed for the comparison of two groups for outcomes normally distributed  with equal variance in the two groups (\cite{Peacock.2012}). While the equal variance hypothesis is convenient to obtain simple formulae, it was possible to extend the method to unequal variance (\cite{Sauzet.2014}) by providing a correction factor. In \cite{Sauzet.2015} the question of the robustness to deviations from normality has been addressed and showed that for small deviations the method worked well. In case of perturbation to the normal distribution (e.g. because of an  excess of patient with hight blood pressure or preterm babies having much lower birthweights) a method based on the skew-normal (\cite{azzalini2005}) distribution has also been proposed in \cite{Sauzet.2015}. 

Because most analysis performed to compare continuous outcomes between groups may involve potentially complex regression models and/or the number of exposed groups may exceed one, we adapted the distributional method to obtain adjusted comparisons of proportions to reflect the results of linear regression models (\cite{sauzet.2016}). 

Some years ago a module for the statical software Stata has been developed by \cite{sauzet.2016b} but so far no package has been available for  \proglang{R} users. The  \proglang{R} Package {\pkg{distdichoR} presented in this article has been developed to cover the application of the distributional methods which have been developed  so far with the addition of an option for the derivation of gamma distributed outcomes which is presented below.  In the following we use examples to illustrate the use of the various commands and options in the package. We first recall how the method works and then illustrate the various commands and options of the  \proglang{R} Package {\pkg{distdichoR} using examples.

\section{Distributional estimates for the comparison of proportions}

\subsection{The normal method}\label{dist}
In this section we review the basic principle of the distributional method as published in \cite{Peacock.2012} and \cite{Sauzet.2014}.

The distributional method is a large sample approximation method for the estimation of proportions and their standard errors assuming a normal distribution for the data.  It is based on the delta method and uses estimates for the mean and variance from the data. We recall here the formulae obtained to compute estimates and standard errors  for proportions, difference in proportions, risk ratios and odds ratios derived from the normal distribution.

Let $\overline{X}_n$ be the sample mean of $n$ independent identically normally distributed random variables $X_i,\  i=1...n$. Let $x_0$ be a real number. The random variable   $p(\overline{X}_n)$ for the proportion of the population with outcome value under the threshold (cutpoint) $x_0$ is defined as 

\begin {equation} p(\overline{X}_n)=\int_{-\infty}^{x_0}f_{N(\overline{X}_n,\sigma^2)}(t)dt\label{eq.prop}\end{equation} where $f_{N(\mu,\sigma^2)}$ is the density function of the normal distribution with mean $\mu$ and variance $\sigma^2$. Hence, it is a function of the sample mean with variance $\sigma^2$. According to the  delta method $p(\overline{X}_n)$ is  asymptotically normally distributed with mean  $p(\overline{x}_n)$ (mean sample estimate) and standard deviation

$$\text{sd}(p(\overline{X}_n))=\frac{s}{\sqrt{n}} f_{N(\overline{x}_n,s^2)}(x_0)$$ so the estimate for the proportion under the quantile $x_o$ is estimated by 
$\int_{-\infty}^{x_0}f_{N(\overline{x}_n,s^2)}(t)dt$ with standard error $\frac{s}{\sqrt{n}} f_{N(\overline{x}_n,s^2)}(x_0)$
where $\overline{x}_n$ is the sample estimate for the mean and $s$ the sample estimate for the standard deviation assumed to be the known standard deviation in the population.

Therefore, for two groups, if the variance is assumed to be the same in both groups,  we obtain estimates for the difference in proportion $d$ as the difference between the estimated proportions with standard error using for the common standard deviation the pooled estimate from the data :

\begin{equation}s_{pooled}=\sqrt{\frac{(n_t-1)s^2_t+(n_c-1)s^2_c)}{(n_t+n_c-2)}}\label{pooled}\end{equation}

\begin{equation}\text{se}(d)=\sqrt{\frac{s_{pooled}^2}{{n_t}}f_{N(\overline{x}_{t,n_t},s_{pooled}^2)}^2(x_0)+\frac{s_{pooled}^2}{{n_c}}f_{N(\overline{x}_{c,n_c},s_{pooled}^2)}^2(x_0)} . \label{eq.diff}\end{equation}

Estimates for the standard error for the log risk ratio $\log(rr)$ is obtained through the function $ h(\overline{X}_n)=\log(p(\overline{X}_n))$. The standard error for the log risk ratio is 

\begin{equation}\text{se}(\log(rr))=\sqrt{\frac{s_{pooled}^2}{{n_t}}\frac{f^2_{N(\overline{x}_{t,n_t}s_{pooled}^2)}(x_o)}{p^2_t}+\frac{s_{pooled}^2}{{n_c}}\frac{f^2_{N(\overline{x}_{c,n_c}s_{pooled}^2)}(x_o)}{p^2_c}} . \label{eq.rr}\end{equation}

Estimates for the standard error for the log odds ratio is obtained through the function $g(\overline{X}_n))=\log(\frac{p(\overline{X}_n)}{1-p(\overline{X}_n)})$. The standard error for the log odds ratio is
\begin{equation}\text{se}(\log(or))=\sqrt{\frac{s_{pooled}^2}{{n_t}}\frac{f^2_{N(\overline{x}_{c,n_c},s_{pooled}^2)}(x_o)}{p_c^2(1-p_c)^2}+\frac{s_{pooled}^2}{{n_c}}\frac{f^2_{N(\overline{x}_{t,n},s_{pooled}^2)}(x_o)}{p_t^2(1-p_t)^2}} . \label{eq.or}\end{equation}

\bigskip

If the hypothesis of equal variance cannot hold, then either providing a known ratio of variance between the two groups or adding a correction factor to the standard error, which otherwise would be underestimated when the variances are not assumed known, is possible. Moreover, this correction factor can also be used to correct the standard errors for  large effects  (see \cite{Sauzet.2014}) as the variability due to using the observed pooled standard deviation  need to be accounted for in the standard error. 

\subsection{For disturbed normal data: using the skew-normal distribution}

For outcomes which are disturbed in their normality,
 the skew-normal distribution \cite{azzalini2005} can be used. 
 This distribution is a generalisation of the normal distribution which works by adding a third parameter $\alpha$  defining the skewness ( $\alpha=0$ gives the normal distribution). We briefly recall how the formula for the standard errors are obtained (\cite{Sauzet.2015}).

 Let $\overline{X}_n$ be the sample mean of $n$ independent identically skew-normal distributed random variables $X_i,\  i=1...n$ with mean $\mu$, variance $\sigma^2$ and skewness parameter $\alpha$. Let $x_0$ be a threshold of interest. The random variable   $p(\overline{X}_n)$ for the proportion of the population with outcome value under the threshold  $x_0$ is defined as 

\begin {equation} p(\overline{X}_n)=\int_{-\infty}^{x_0}2\frac{e^{\frac{-1}{2w^2}(t-(\overline{X}_n+\alpha'))^2}}{\sqrt{2\pi w^2}}\left(\int_{-\infty}^{\alpha(t-(\overline{X}_n+\alpha'))/w}\frac{e^{\frac{-1}{2}r^2}}{\sqrt{2\pi}}dr\right)dt\end{equation}

where $\alpha'=\mu-w\mu_z$ and $w^2=\sigma^2/(1-\mu_z^2)$ with $ \mu_z^2=\frac{2}{\pi}\frac{\alpha^2}{1+\alpha^2}$ (see \cite{azzalini2005}).

From the delta method we obtain that $  p(\overline{X}_n)$ is approximately normally distributed with standard deviation $$\frac{w^2}{\sqrt{n}}\left(1-\mu_z^2\right)p'(\mu)^2.$$

After the derivation of $p'(\mu)$ the formula for the standard error for the difference in proportions $d$, log risk ration $log(rr)$ and log odds ration $log(or)$ obtained are the following:
$$se(d)^2=\frac{w_1^2}{\sqrt{n_1}}\left(1-\mu_z^2\right)\left(\frac{2e^{\frac{-1}{2w_1^2}(x_0-(\mu_1+\alpha_1'))^2}}{\sqrt{2\pi w_1^2}}
\Phi(\alpha\frac{x_0-(\mu_1-\alpha_1')}{w_1}\right)^2    +$$$$\frac{w_2^2}{\sqrt{n_2}}\left(1-\mu_z^2\right)\left(\frac{2e^{\frac{-1}{2w_2^2}(x_0-(\mu_2+\alpha_2'))^2}}{\sqrt{2\pi w_2^2}}
\Phi(\alpha\frac{x_0-(\mu_2-\alpha_2')}{w_2}  \right )^2      $$

$$se(log(rr))^2=\frac{1}{p_1^2}\frac{w_1^2}{\sqrt{n_1}}\left(1-\mu_z^2\right)\left(\frac{2e^{\frac{-1}{2w_1^2}(x_0-(\mu_1+\alpha_1'))^2}}{\sqrt{2\pi w_1^2}}
\Phi(\alpha\frac{x_0-(\mu_1-\alpha_1')}{w_1}\right)^2    +$$$$\frac{1}{p_2^2}\frac{w_2^2}{\sqrt{n_2}}\left(1-\mu_z^2\right)\left(\frac{2e^{\frac{-1}{2w_2^2}(x_0-(\mu_2+\alpha_2'))^2}}{\sqrt{2\pi w_2^2}}
\Phi(\alpha\frac{x_0-(\mu_2-\alpha_2')}{w_2}  \right )^2      $$

$$se(log(or))^2=\frac{1}{(p_1(1-p_1))^2}\frac{w_1^2}{\sqrt{n_1}}\left(1-\mu_z^2\right)\left(\frac{2e^{\frac{-1}{2w_1^2}(x_0-(\mu_1+\alpha_1'))^2}}{\sqrt{2\pi w_1^2}}
\Phi(\alpha\frac{x_0-(\mu_1-\alpha_1')}{w_1}\right)^2    +$$$$\frac{1}{(p_2(1-p_2))^2}\frac{w_2^2}{\sqrt{n_2}}\left(1-\mu_z^2\right)\left(\frac{2e^{\frac{-1}{2w_2^2}(x_0-(\mu_2+\alpha_2'))^2}}{\sqrt{2\pi w_2^2}}
\Phi(\alpha\frac{x_0-(\mu_2-\alpha_2')}{w_2}  \right )^2 .     $$

\subsection{Using the gamma distribution}

The \proglang{R} package \code{distdichoR} also has an option for the dichotomisation of gamma distributed outcomes for which we provide the derivation for the standard error following the same principle as above.  Here we assume that the shape parameter $\alpha$ is known. We want to estimate difference (or ratio) of the proportion of the population below (or above) a given cut-point $x_0$. For this we need to derive a formula for the standard error of a proportion assuming a gamma distribution using the delta method as for the normal case.

If $X_1,\dots X_n$ are $n$ gamma distributed independent variables with shape parameter $\alpha$ and scale parameter $\beta$, then the distribution mean $\mu$ is $$\mu=\frac{\alpha}{\bar\beta}$$ such that the random variable $p(\bar{X_n})$ is given by

$$p(\bar{X_n})=\frac{1}{\Gamma(\alpha)}\int_0^{x_0}\left(\frac{\alpha}{\bar{X_n}}\right)^\alpha x^{\alpha-1}e^{-\frac{\alpha x}{\bar{X_n}}}dx.$$

The variance of $p(\bar{X_n})$ is obtained by differentiating $p(\bar{X_n})$ with respect to $\bar{X_n}$

$$p'(\bar{X_n})=\frac{1}{\Gamma(\alpha)}\int_0^{x_0}x^{\alpha-1}\frac{d\left[\left(\frac{\alpha}{\bar{X_n}}\right)^\alpha e^{-\frac{\alpha x}{\bar{X_n}}}\right]}{d\bar{X_n}}dx.$$

An integration by part using $$A(\bar{X_n})=\frac{d\left[\left(\frac{\alpha}{\bar{X_n}}\right)^\alpha e^{-\frac{\alpha x}{\bar{X_n}}}\right]}{d\bar{X_n}} \text{, \  \   } B(\bar{X_n})=\left(\frac{\alpha}{\bar{X_n}}\right)^\alpha \text{ and \ \  } C(\bar{X_n})=e^{-\frac{\alpha x}{\bar{X_n}}}$$

provides


$$A(\bar{X_n})=B(\bar{X_n})C'(\bar{X_n})+B'(\bar{X_n})C(\bar{X_n})=\left(\frac{\alpha}{\bar{X_n}}\right)^\alpha\left(\frac{\alpha x}{\bar{X_n}^2}\right)e^{-\frac{\alpha x}{\bar{X_n}}}-\left(\frac{\alpha^{\alpha+1}}{\bar{X_n}^{\alpha+1}}\right)e^{-\frac{\alpha x}{\bar{X_n}}}$$
$$=\left(\frac{\alpha^{\alpha+1} x}{\bar{X_n}^{\alpha+2}}\right)e^{-\frac{\alpha x}{\bar{X_n}}}-\left(\frac{\alpha^{\alpha+1}}{\bar{X_n}^{\alpha+1}}\right)e^{-\frac{\alpha x}{\bar{X_n}}}$$

such that the derivative of $p(\bar{X_n})$ is 
$$p'(\bar{X_n})=\frac{1}{\Gamma(\alpha)}\int_0^{x_0}x^{\alpha-1}\left[\left(\frac{\alpha^{\alpha+1} x}{\bar{X_n}^{\alpha+2}}\right)e^{-\frac{\alpha x}{\bar{X_n}}}-\left(\frac{\alpha^{\alpha+1}}{\bar{X_n}^{\alpha+1}}\right)e^{-\frac{\alpha x}{\bar{X_n}}}\right]dx$$
$$=\frac{1}{\Gamma(\alpha)}\int_0^{x_0}x^{\alpha-1}\left(\frac{\alpha^{\alpha+1} x}{\bar{X_n}^{\alpha+2}}\right)e^{-\frac{\alpha x}{\bar{X_n}}}dx-\frac{1}{\Gamma(\alpha)}\int_0^{x_0}x^{\alpha-1}\left(\frac{\alpha^{\alpha+1}}{\bar{X_n}^{\alpha+1}}\right)e^{-\frac{\alpha x}{\bar{X_n}}}dx$$


such that
$$p'(\bar{X_n})=-\frac{1}{\Gamma(\alpha)}\frac{\alpha^{\alpha}}{\bar{X_n}^{\alpha+1}}x_0^\alpha e^{-\frac{\alpha x_0}{\bar{X_n}}}+\frac{\alpha}{\bar{X_n}}p(\bar{X_n})-\frac{\alpha}{\bar{X_n}}p(\bar{X_n})=-\frac{1}{\Gamma(\alpha)}\frac{\alpha^{\alpha}}{\bar{X_n}^{\alpha+1}}x_0^\alpha e^{-\frac{\alpha x_0}{\bar{X_n}}} .$$

Which give for the variance of $p(\bar{X_n})$:
$$\sigma^2 p'(\bar{X_n})^2/n=\frac{\bar{X_n}^2}{\alpha}\left(\frac{1}{\Gamma(\alpha)}\frac{\alpha^{\alpha}}{\bar{X_n}^{\alpha+1}}x_0^\alpha e^{-\frac{\alpha x_0}{\bar{X_n}}}\right)^2/n=\frac{1}{n\alpha}\left(\frac{1}{\Gamma(\alpha)}\frac{\alpha^{\alpha}}{\bar{X_n}^{\alpha}}x_0^\alpha e^{-\frac{\alpha x_0}{\bar{X_n}}}\right)^2 .$$

This method has been checked to provide reliable results for difference in proportions using simulation (data not shown).

\subsection{Adjusted comparisons}

Distributional estimates can also be obtained to describe an adjusted difference in means, i.e. following a linear regression model. A regression model of the form 
$$Y_i=\beta_0+\beta_{r_i }+\beta X_i+\epsilon_i$$
where $Y$ is a random variable and $\epsilon_i$ is the error term for observation $i$ following a normal distribution with a mean of 0 and variance $\sigma_e^2$, is the starting point. An exposure is defined by a categorical variable $R$ with $k+1$ levels, e.g. not smoking during pregnancy, smoking regularly, smoking occasionally. We recall how the distributional method can be used in the context of a regression model (see also \cite{Sauzet2016}).

Then, using the marginal  means $E(Y|R=r)$ for the $k+1$ levels of exposures, we obtain $k+1$ adjusted distributional probabilities for each level of the exposure $r=0,1,..,k$, such that in the normal case
$$p_r=P(Y<a|R=r)=P(\epsilon +E(Y|R=r)<a)=$$$$\Phi\left(\frac{a-E(Y|R=r)}{\sigma_e}\right)$$
for a linear regression.

The method can be generalised to mixed models, for example to a simple random intercept model with two levels
$$Y_i=\beta_0+\beta_{r_i }+\beta X_i+ \mu_i+\epsilon_i$$

where $\beta$ is a vector of fixed effects and $\mu_i$ a  random element with mean zero and a variance $\sigma_r^2$ and the error term $\epsilon_i$ with variance $\sigma_e^2$. Then:

$$p_r=P(Y<a|R=r)=P(\mu+\epsilon+E(Y|R=r)<a)=$$$$\Phi\left(\frac{a-E(Y|R=r)}{\sqrt{\sigma_e^2+\sigma_r^2}} \right).$$

The standard errors are obtained as seen in section \ref{dist}.

\section[The normal case in details]{The normal case in details: the \code{distdicho} command}

Because the distributional method is a complement to a comparison of means,  the \code{distdicho} and  its accompaning \code{distdichoi} commands first return the results of a t-test followed by a table containing the relevant information for each group and the distributional estimates for difference in proportions, risk ratio and odds ratio, their standard error and a 95\% confidence interval. The 95\% confidence interval is based on the assumption of a normal distribution of the estimate. The \code{distdichoi} command works without any individual data by requiring  sample sizes means and standard deviations only.
\medskip

The \code{distdicho} command can be used under the assumption that the outcome is normally distributed and allow for a range of assumptions regarding the variance and large effects. 

\medskip

In the sequel the various options will be discussed through worked examples.

\subsection{Examples}

 We use data from  the St George's Birthweight Study (\cite{peacock1995}) containing the outcome variables birthweight (BW), body-mass index (BMI) and gestational age (GA) for which a dichotomisation is common. We divived four example datasets \code{bwsmoke, bmi, bmi2}, and \code{bwsmokecompl}.  The options of the command  \code{distdicho} are reviewed below for  various group comparisons including the smoking status during pregnancy yes (1)/no (0), parity primipari (0) /multipari (1) (first or subsequent pregnancies), employed (1)/unemployed (2).

{\subsubsection{Example 1}
This first dataset contains the birthweight of 1458  term live births  (gestational age ({\it gest}) greater or equal to 37 weeks). Live term birth are known to be normally distributed (\cite{Wilcox.2001}), but we can check that it is the case here by plotting the outcomes in the two groups of smoking and non smoking mothers (see Fig. \ref{Fig:bwsmoke.birthwt}). Low-birthweight babies are defined as those with a birthweight below 2500g (\cite{who}). 

\begin{figure}[h!]
  \centering
	\includegraphics[width=0.8\textwidth]{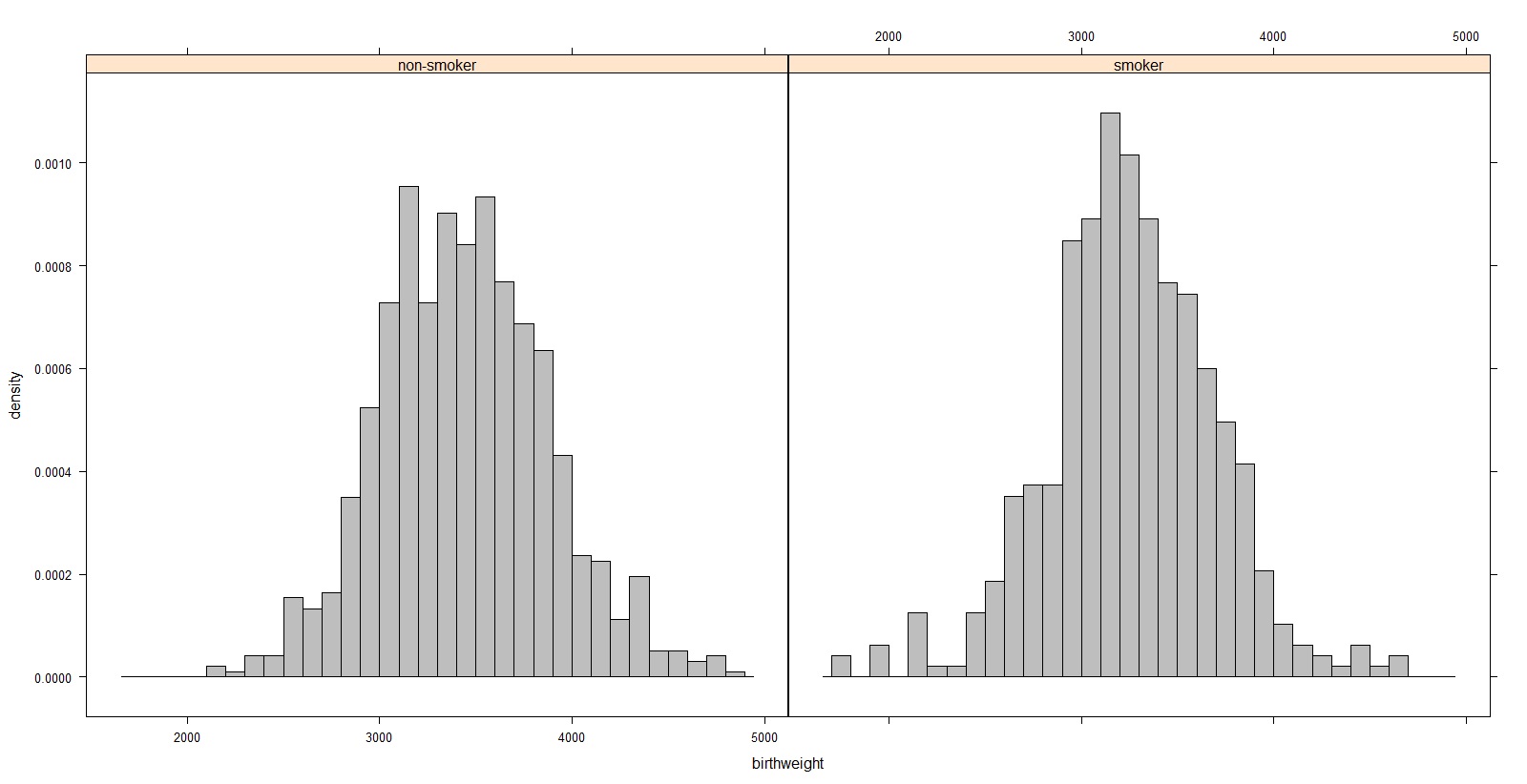}
	\caption{Histogram of birth weights by smokers (1) and non-smokers (0)}
	\label{Fig:bwsmoke.birthwt}
\end{figure}

We see no evidence of unequal variance between smokers and non-smokers, therefore we can apply the simplest form of the distributional method using the cut-point 2500g to obtain the comparison of proportion of babies with low birthweight.

\begin{knitrout}\footnotesize
\definecolor{shadecolor}{rgb}{0.969, 0.969, 0.969}\color{fgcolor}\begin{kframe}
\begin{alltt}
\hlkwd{library}\hlstd{(distdichoR)}
\hlkwd{distdicho}\hlstd{(birthwt} \hlopt{~} \hlstd{smoke,} \hlkwc{cp} \hlstd{=} \hlnum{2500}\hlstd{,} \hlkwc{data} \hlstd{= bwsmoke,} \hlkwc{exposed} \hlstd{=} \hlstr{'smoker'}\hlstd{)}
\end{alltt}
\begin{verbatim}
## ======================================================
## ===              t-Test                            ===
## ======================================================
## 
## 	Two Sample t-test
## 
## data:  x and y
## t = -7.6418, df = 1456, p-value = 3.864e-14
## alternative hypothesis: true difference in means is not equal to 0
## 95 percent confidence interval:
##  -233.4477 -138.0791
## sample estimates:
## mean of x mean of y 
##  3266.965  3452.728 
## 
## ======================================================
## ===              Distributional method             ===
## ======================================================
## Distributional estimates for the comparison of proportions below the cut-point 2500 
## Standard error computed under the hypothesis that the ratio of variances is equal to 1 
## 
##       Group Obs     Mean  Std.Dev Dist.prop.
##      smoker 483 3266.965 437.7330 0.03958289
##  non-smoker 975 3452.728 436.4585 0.01460092
## 
## ------------------------------------------------------
##        Stat   Estimate     Std.Err   CI.lower   CI.upper
##  Diff. prop 0.02498197 0.004064361 0.01701597 0.03294797
##  Risk ratio 2.71098543 0.349646388 2.11190092 3.48001270
##  Odds ratio 2.78150245 0.369993261 2.15034801 3.59790874
## 
## ------------------------------------------------------
## * 95 percent confidence interval
## * confidence interval calculated using distributional standard error 
## 
## ------------------------------------------------------
\end{verbatim}
\end{kframe}
\end{knitrout}

In this dataset,  mothers who smoke have on average babies weighting 186 g less than mothers who do not smoke during pregnancy. This difference, assuming the normality of the outcome,  corresponds to a difference in proportions of low birthweight babies  of almost 2.5 percentage points (difference in proportions: 0.025) between smoking and non-smoking mothers. The confidence interval for this difference, [0.017, 0.033], reflects the  precision  of the difference in means.

\subsubsection{Example 2}

A transformation may enable the use of \code{distdicho} in the case of a skewed outcome. Inverse BMI is reasonably normally distributed therefore we can use the distributional method to compare the proportion of obese mothers (BMI$\geq$30) at the beginning of pregnancy between primi and multipari mothers. The proportion of interest is in the upper tail of the distribution of BMIs but it is in the lower tail of the inverse BMI because inverse is a decreasing function on positive values. The transformed cut-point is equal to $1/30\simeq 0.033$.

\begin{knitrout}\footnotesize
\definecolor{shadecolor}{rgb}{0.969, 0.969, 0.969}\color{fgcolor}\begin{kframe}
\begin{alltt}
\hlkwd{library}\hlstd{(distdichoR)}
\hlkwd{distdicho}\hlstd{(inv_bmi} \hlopt{~} \hlstd{parity ,} \hlkwc{cp} \hlstd{=} \hlnum{0.033}\hlstd{,} \hlkwc{data} \hlstd{= bmi,} \hlkwc{exposed} \hlstd{=} \hlstr{'primi'}\hlstd{)}
\end{alltt}
\begin{verbatim}
## ======================================================
## ===              t-Test                            ===
## ======================================================
## 
## 	Two Sample t-test
## 
## data:  x and y
## t = 5.0304, df = 1779, p-value = 5.387e-07
## alternative hypothesis: true difference in means is not equal to 0
## 95 percent confidence interval:
##  0.0008804219 0.0020056651
## sample estimates:
##  mean of x  mean of y 
## 0.04439543 0.04295239 
## 
## ======================================================
## ===              Distributional method             ===
## ======================================================
## Distributional estimates for the comparison of proportions below the cut-point 0.033 
## Standard error computed under the hypothesis that the ratio of variances is equal to 1 
## 
##  Group Obs       Mean     Std.Dev Dist.prop.
##  primi 891 0.04439543 0.005884324 0.02987782
##  multi 890 0.04295239 0.006217391 0.05006816
## 
## ------------------------------------------------------
##        Stat    Estimate     Std.Err    CI.lower    CI.upper
##  Diff. prop -0.02019034 0.004139886 -0.02830437 -0.01207631
##  Risk ratio  0.59674290 0.061808630  0.48788597  0.72988794
##  Odds ratio  0.58432339 0.063031215  0.47382541  0.72058994
## 
## ------------------------------------------------------
## * 95 percent confidence interval
## * confidence interval calculated using distributional standard error 
## 
## ------------------------------------------------------
\end{verbatim}
\end{kframe}
\end{knitrout}

We obtain that the difference in proportions of obesity among multipari mothers (non-exposed) is 2 percentage points higher than among primipari mothers (exposed). We also can see that the risk of obesity is 1.68 times higher among multipari mothers than among primipari, and the odds of obesity are 1.71 times higher. It is to be noted that an usual drawback of using transformation is that the results are difficult to interpret in the original scale, but the comparison of proportions can be interpreted equally, no matter what scale is used for the analysis.

\subsubsection{Example 3}

We now compare the proportion of obese mothers between those who are employed and those who are not. However, the standard deviations cannot be assumed to be equal in both groups (see \cite{Sauzet.2014}).

 If we fail to have any theoretical base to provide a known variance ratio, we use a correction factor with the option \code{R=0}. An equivalent alternative would be to use the option \code{uneq}.

\begin{knitrout}\footnotesize
\definecolor{shadecolor}{rgb}{0.969, 0.969, 0.969}\color{fgcolor}\begin{kframe}
\begin{alltt}
\hlkwd{library}\hlstd{(distdichoR)}
\hlkwd{distdicho}\hlstd{(inv_bmi} \hlopt{~} \hlstd{employ,} \hlkwc{cp} \hlstd{=} \hlnum{0.033}\hlstd{,} \hlkwc{R} \hlstd{=} \hlnum{0}\hlstd{,} \hlkwc{data} \hlstd{= bmi2,} \hlkwc{exposed} \hlstd{=} \hlstr{'employed'}\hlstd{)}
\end{alltt}
\begin{verbatim}
## ======================================================
## ===              t-Test                            ===
## ======================================================
## 
## 	Welch Two Sample t-test
## 
## data:  x and y
## t = 1.5199, df = 1417.4, p-value = 0.1288
## alternative hypothesis: true difference in means is not equal to 0
## 95 percent confidence interval:
##  -0.0001371404  0.0010807694
## sample estimates:
##  mean of x  mean of y 
## 0.04385758 0.04338577 
## 
## ======================================================
## ===              Distributional method             ===
## ======================================================
## Distributional estimates for the comparison of proportions below the cut-point 0.033 
## Standard error computed with correction for unknown variance ratio
## 
##       Group Obs       Mean     Std.Dev Dist.prop.
##    employed 851 0.04385758 0.005646543 0.02724804
##  unemployed 709 0.04338577 0.006462314 0.05401313
## 
## ------------------------------------------------------
##        Stat    Estimate     Std.Err    CI.lower    CI.upper
##  Diff. prop -0.02676509 0.006878875 -0.04024744 -0.01328275
##  Risk ratio  0.50447062 0.097880039  0.34837479  0.73050811
##  Odds ratio  0.49059020 0.097648188  0.33573088  0.71687999
## 
## ------------------------------------------------------
## * 95 percent confidence interval
## * confidence interval calculated using distributional standard error 
## 
## ------------------------------------------------------
\end{verbatim}
\end{kframe}
\end{knitrout}

The distributional difference in proportions of obesity among unemployed mothers is 2.7
percentage points higher than among employed mothers. It also shows that the risk of
obesity (RR) is almost twice as high among unemployed versus employed mothers.

\subsubsection{Example 4}

If we have reasons to assume that the ratio of variance unemployed/employed is 1.3, then the comparison of proportions are obtained without correction using this time \code{R=1.3}.

\begin{knitrout}\footnotesize
\definecolor{shadecolor}{rgb}{0.969, 0.969, 0.969}\color{fgcolor}\begin{kframe}
\begin{alltt}
\hlkwd{library}\hlstd{(distdichoR)}
\hlkwd{distdicho}\hlstd{(inv_bmi} \hlopt{~} \hlstd{employ,} \hlkwc{cp} \hlstd{=} \hlnum{0.033}\hlstd{,} \hlkwc{R} \hlstd{=} \hlnum{1.3}\hlstd{,} \hlkwc{data} \hlstd{= bmi2,} \hlkwc{exposed} \hlstd{=} \hlstr{'employed'}\hlstd{)}
\end{alltt}
\begin{verbatim}
## ======================================================
## ===              t-Test                            ===
## ======================================================
## 
## 	Welch Two Sample t-test
## 
## data:  x and y
## t = 1.5199, df = 1417.4, p-value = 0.1288
## alternative hypothesis: true difference in means is not equal to 0
## 95 percent confidence interval:
##  -0.0001371404  0.0010807694
## sample estimates:
##  mean of x  mean of y 
## 0.04385758 0.04338577 
## 
## ======================================================
## ===              Distributional method             ===
## ======================================================
## Distributional estimates for the comparison of proportions below the cut-point 0.033 
## Standard error computed under the hypothesis that the ratio of variances is equal to 1.3 
## 
##       Group Obs       Mean     Std.Dev Dist.prop.
##    employed 851 0.04385758 0.005646543 0.04705998
##  unemployed 709 0.04338577 0.006462314 0.03394227
## 
## ------------------------------------------------------
##        Stat   Estimate     Std.Err    CI.lower   CI.upper
##  Diff. prop 0.01311771 0.004399089 0.004495652 0.02173976
##  Risk ratio 1.38647115 0.153726591 1.117850858 1.71964107
##  Odds ratio 1.40555663 0.162387884 1.123227227 1.75885111
## 
## ------------------------------------------------------
## * 95 percent confidence interval
## * confidence interval calculated using distributional standard error 
## 
## ------------------------------------------------------
\end{verbatim}
\end{kframe}
\end{knitrout}

Here we used the observed  ratio of variances as the known ratio. Therefore, the estimates obtained in Examples 3 and 4 are the same. However, by assuming that we not know the "true" ratio, we allowed for more uncertainty, thus the standard errors in Example 3 are larger than in Example 4.

\section[The distdichogen commands]{A choice of distribution: the \code{distdichogen}  command }

The \code{distdichogen} command has the same syntax as the \code{distdicho} command but allows a wider range of distribution for the outcome: normal, skew normal or gamma.   Each work under the assumption that the shape of the distribution is the same in both groups and that the parameter for the shape is known.

\subsection{Examples}

\subsubsection{Example 5}

In this example we show that  assuming a normal distribution or a skew-normal distribution lead to similar results when the data is approximatively normally distributed. We reproduce Example 1  with babies birthweights using the command \code{distdichogen} with \code{dist=sk\_normal} instead of \code{distdicho}.

\begin{knitrout}\footnotesize
\definecolor{shadecolor}{rgb}{0.969, 0.969, 0.969}\color{fgcolor}\begin{kframe}
\begin{alltt}
\hlkwd{library}\hlstd{(distdichoR)}
\hlkwd{distdichogen}\hlstd{(birthwt} \hlopt{~} \hlstd{smoke,} \hlkwc{cp} \hlstd{=} \hlnum{2500}\hlstd{,} \hlkwc{data} \hlstd{= bwsmoke,} \hlkwc{exposed} \hlstd{=} \hlstr{'smoker'}\hlstd{,}
             \hlkwc{dist} \hlstd{=} \hlstr{'sk_normal'}\hlstd{)}
\end{alltt}
\begin{verbatim}
## ======================================================
## ===              t-Test                            ===
## ======================================================
## 
## 	Two Sample t-test
## 
## data:  x and y
## t = -7.6418, df = 1456, p-value = 3.864e-14
## alternative hypothesis: true difference in means is not equal to 0
## 95 percent confidence interval:
##  -233.4477 -138.0791
## sample estimates:
## mean of x mean of y 
##  3266.965  3452.728 
## 
## ======================================================
## ===              Distributional method             ===
## ======================================================
## Distributional estimates for the comparison of proportions below the cut-point 2500 
## Standard error computed under the hypothesis that the ratio of variances is equal to 1 
## 
## Alpha: 0.8668926 
## 
##       Group Obs     Mean  Std.Dev Dist.prop.
##      smoker 483 3266.965 437.7330 0.03656510
##  non-smoker 975 3452.728 436.4585 0.01249528
## 
## ------------------------------------------------------
##        Stat   Estimate     Std.Err   CI.lower  CI.upper
##  Diff. prop 0.02406982 0.004142819 0.01595005 0.0321896
##  Risk ratio 2.92631339 0.487284177 2.12513449 4.0295380
##  Odds ratio 2.99942247 0.511518697 2.16213025 4.1609589
## 
## ------------------------------------------------------
## * 95 percent confidence interval
## * confidence interval calculated using distributional standard error 
## 
## ------------------------------------------------------
\end{verbatim}
\end{kframe}
\end{knitrout}

The estimated skew parameter $\alpha$ is not exactly one but the estimates as well as their standard errors for the difference in proportion are almost unchanged in comparison with Example 1. The distributional method has been shown to be robust to small variations to normality  \cite{Sauzet.2015} which explain the results observed here.
However, because the estimated proportions for each group vary between Example 1 and
Example 5, the RR and OR also vary between these two examples.

\subsubsection{Example 6}

In Example 2 we used a transformation to obtain a normally distributed outcome. We use the same data to compare the skew-normal approach to the transformation approach. Note that now the proportions of interest (obesity) is in the upper tail of the distribution.

\begin{knitrout}\footnotesize
\definecolor{shadecolor}{rgb}{0.969, 0.969, 0.969}\color{fgcolor}\begin{kframe}
\begin{alltt}
\hlkwd{library}\hlstd{(distdichoR)}
\hlkwd{distdichogen}\hlstd{(bmi} \hlopt{~} \hlstd{group_par,} \hlkwc{cp} \hlstd{=} \hlnum{30}\hlstd{,} \hlkwc{data} \hlstd{= bmi,} \hlkwc{exposed} \hlstd{=} \hlstr{'1'}\hlstd{,} \hlkwc{tail} \hlstd{=} \hlstr{'upper'}\hlstd{,}
             \hlkwc{dist} \hlstd{=} \hlstr{'sk_normal'}\hlstd{)}
\end{alltt}
\begin{verbatim}
## ======================================================
## ===              t-Test                            ===
## ======================================================
## 
## 	Two Sample t-test
## 
## data:  x and y
## t = 4.9986, df = 1779, p-value = 6.341e-07
## alternative hypothesis: true difference in means is not equal to 0
## 95 percent confidence interval:
##  0.5345447 1.2248856
## sample estimates:
## mean of x mean of y 
##  23.84148  22.96176 
## 
## ======================================================
## ===              Distributional method             ===
## ======================================================
## Distributional estimates for the comparison of proportions above the cut-point 30 
## Standard error computed under the hypothesis that the ratio of variances is equal to 1 
## 
## Alpha: 4.119313 
## 
##  Group Obs     Mean  Std.Dev Dist.prop.
##      1 890 23.84148 4.012678 0.06835550
##      0 891 22.96176 3.388547 0.04858033
## 
## ------------------------------------------------------
##        Stat   Estimate     Std.Err   CI.lower   CI.upper
##  Diff. prop 0.01977517 0.004015664 0.01190461 0.02764573
##  Risk ratio 1.40706129 0.096539107 1.23059774 1.60882913
##  Odds ratio 1.43692770 0.104711089 1.24638216 1.65660364
## 
## ------------------------------------------------------
## * 95 percent confidence interval
## * confidence interval calculated using distributional standard error 
## 
## ------------------------------------------------------
\end{verbatim}
\end{kframe}
\end{knitrout}

The estimates and standard errors obtained here and in Example 2 are very close because the transformation used in example 2 was quite successful in providing an approximatively normal distribution.
 We still have, for example, a difference of about 2 percentage points in
proportions of obesity between multipari and primipari mothers.

\subsubsection{Example 7}

The Apgar score \cite{apgar2015proposal} aim is to make a quick decision about the needs of a newborn for medical help. It takes values between 0 and 10 so that a baby with an Apgar score 0 needs urgent medical attention and with values of 7 or above the baby is normal. The transformed Apgar score \code{10-Apgar}, which is 10 minus the Apgar score, can be seen to be approximatively gamma distributed (but the score is discrete). We use the \code{distdichogen} command with option \code{gamma} to check if there is a difference in proportion of babies with Apgar score below 7 (or on the transformed scale: above 3) between smoking and non-smoking mothers.

\begin{figure}[h!]
  \centering
	\includegraphics[width=0.8\textwidth]{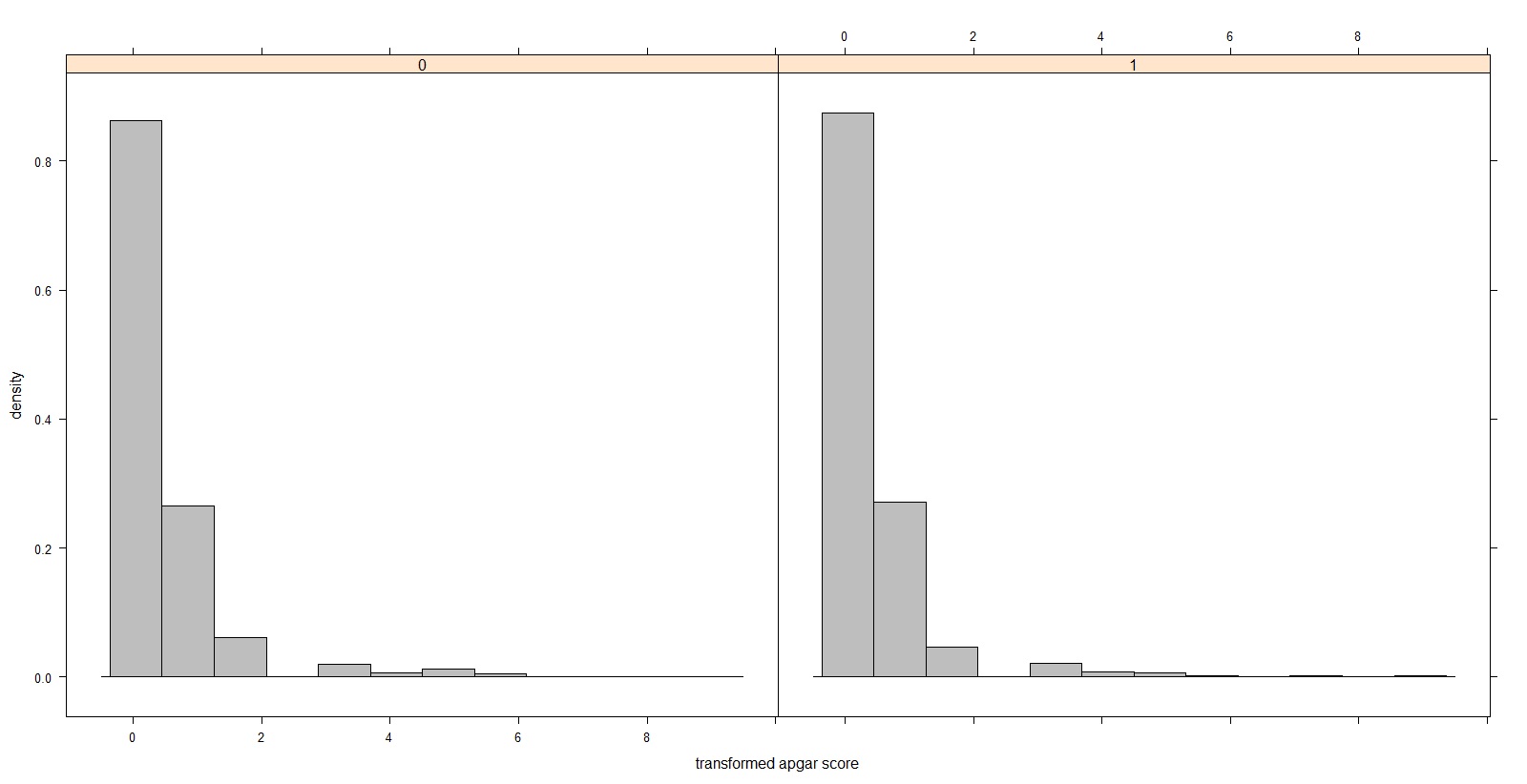}
	\caption{Histogram of the transformed Apgar score by smokers (1) and non-smokers (0)}
	\label{Fig:bwsmokecompl.apgar}
\end{figure}

\begin{knitrout}\footnotesize
\definecolor{shadecolor}{rgb}{0.969, 0.969, 0.969}\color{fgcolor}\begin{kframe}
\begin{alltt}
\hlkwd{library}\hlstd{(distdichoR)}
\hlkwd{distdichogen}\hlstd{(apgar_10} \hlopt{~} \hlstd{smoke,} \hlkwc{cp} \hlstd{=} \hlnum{3}\hlstd{,} \hlkwc{data} \hlstd{= bwsmokecompl,} \hlkwc{exposed} \hlstd{=} \hlstr{'1'}\hlstd{,} \hlkwc{tail} \hlstd{=} \hlstr{'upper'}\hlstd{,}
             \hlkwc{dist} \hlstd{=} \hlstr{'gamma'}\hlstd{)}
\end{alltt}
\begin{verbatim}
## ======================================================
## ===              t-Test                            ===
## ======================================================
## 
## 	Two Sample t-test
## 
## data:  x and y
## t = -0.66013, df = 1903, p-value = 0.5092
## alternative hypothesis: true difference in means is not equal to 0
## 95 percent confidence interval:
##  -0.11786668  0.05850183
## sample estimates:
## mean of x mean of y 
## 0.4331210 0.4628034 
## 
## ======================================================
## ===              Distributional method             ===
## ======================================================
## Distributional estimates for the comparison of proportions above the cut-point 3 
## Standard error computed under the hypothesis that the ratio of variances is equal to 1 
## 
## Alpha: 0.2371702 
## 
##  Group  Obs      Mean   Std.Dev Dist.prop.
##      1  628 0.4331210 0.9108517 0.02579994
##      0 1277 0.4628034 0.9282585 0.02973904
## 
## ------------------------------------------------------
##        Stat     Estimate     Std.Err   CI.lower    CI.upper
##  Diff. prop -0.003939105 0.005862453 -0.0154293 0.007551093
##  Risk ratio  0.867544322 0.030906529  0.8090887 0.930223242
##  Odds ratio  0.864036471 0.031208707  0.8050391 0.927357475
## 
## ------------------------------------------------------
## * 95 percent confidence interval
## * confidence interval calculated using distributional standard error 
## 
## ------------------------------------------------------
\end{verbatim}
\end{kframe}
\end{knitrout}

While the difference in proportions are a good reflection of the difference in means with no significant difference between the two groups of mothers, the OR and RR are not. This is probably due to a lower robustness of OR and RR to deviation for the distributional assumption than to the difference in proportions.

\section[The regdistdicho command]{With a linear regression: the \code{regdistdicho} command}

The regdistdicho command provides distributional estimates for a comparison of proportion adjusted for independent variables. It can be used with the commands \code{lm}
for a linear model or \code{lme} a linear multilevel model (\proglang{R} package \code{nlme} \cite{nlme}).

\subsection{Examples}

The two examples below are based on the dataset \code{bwsmokecompl} in which all live birth are used. The data has been slightly modified to obtain a third category for smoking which includes mothers which did not provide their smoking status (154 observations coded ``2''). Moreover, to introduce a hierarchical structure, we added 20\% twins (see Example 9 below) which makes a total of 2083 observations, where 351 have been created.

\subsubsection{Example 8}

Example 7 is revisited again, but this time with all live birth. Gestational age explains the skweness in the live birth birthweight, so we going to adjust for that in order to have normally distributed residuals. We would like the estimate of proportion comparison adjusted for gestational age and the effect of smoking (or not) including for mothers who did not provide their smoking status, thus adding a second difference in proportion.

\begin{knitrout}\footnotesize
\definecolor{shadecolor}{rgb}{0.969, 0.969, 0.969}\color{fgcolor}\begin{kframe}
\begin{alltt}
\hlkwd{library}\hlstd{(distdichoR)}
\hlstd{mod_smoke} \hlkwb{<-} \hlkwd{lm}\hlstd{(birthwt} \hlopt{~} \hlstd{smoke2} \hlopt{+} \hlstd{gest,} \hlkwc{data} \hlstd{= bwsmokecompl)}
\hlkwd{summary}\hlstd{(mod_smoke)}
\end{alltt}
\begin{verbatim}
## 
## Call:
## lm(formula = birthwt ~ smoke2 + gest, data = bwsmokecompl)
## 
## Residuals:
##      Min       1Q   Median       3Q      Max 
## -1437.58  -268.91    -5.48   271.96  1433.38 
## 
## Coefficients:
##              Estimate Std. Error t value Pr(>|t|)    
## (Intercept) -2984.076    146.624 -20.352  < 2e-16 ***
## smoke21      -153.412     20.564  -7.460 1.26e-13 ***
## smoke22      -121.755     33.152  -3.673 0.000246 ***
## gest          161.145      3.751  42.961  < 2e-16 ***
## ---
## Signif. codes:  0 '***' 0.001 '**' 0.01 '*' 0.05 '.' 0.1 ' ' 1
## 
## Residual standard error: 420.2 on 2079 degrees of freedom
##   (23 observations deleted due to missingness)
## Multiple R-squared:  0.4798,	Adjusted R-squared:  0.4791 
## F-statistic: 639.2 on 3 and 2079 DF,  p-value: < 2.2e-16
\end{verbatim}
\begin{alltt}
\hlkwd{regdistdicho}\hlstd{(}\hlkwc{mod} \hlstd{= mod_smoke,} \hlkwc{group_var} \hlstd{=} \hlstr{'smoke2'}\hlstd{,} \hlkwc{cp} \hlstd{=} \hlnum{2500}\hlstd{)}
\end{alltt}
\begin{verbatim}
## [[1]]
## ======================================================
## ===              Distributional method             ===
## ======================================================
## Distributional estimates for the comparison of proportions below the cut-point 2500 
## Standard error computed under the hypothesis that the ratio of variances is equal to 1 
## 
##  Group  Obs     Mean  Std.Dev Dist.prop.
##      1  620 3135.952 420.2215 0.06509255
##      0 1279 3289.364 420.2215 0.03015999
## 
## ------------------------------------------------------
##        Stat   Estimate    Std.Err   CI.lower   CI.upper
##  Diff. prop 0.03493257 0.00544421 0.02426211 0.04560302
##  Risk ratio 2.15824219 0.21907813 1.77154425 2.62934969
##  Odds ratio 2.23888433 0.23986613 1.81804992 2.75713168
## 
## ------------------------------------------------------
## * 95 percent confidence interval
## ------------------------------------------------------
## [[2]]
## ======================================================
## ===              Distributional method             ===
## ======================================================
## Distributional estimates for the comparison of proportions below the cut-point 2500 
## Standard error computed under the hypothesis that the ratio of variances is equal to 1 
## 
##  Group  Obs     Mean  Std.Dev Dist.prop.
##      2  184 3167.609 420.2215 0.05606327
##      0 1279 3289.364 420.2215 0.03015999
## 
## ------------------------------------------------------
##        Stat   Estimate     Std.Err    CI.lower   CI.upper
##  Diff. prop 0.02590329 0.008542252 0.009160782 0.04264579
##  Risk ratio 1.85886272 0.306059211 1.354607951 2.55082704
##  Odds ratio 1.90987318 0.332516191 1.367729797 2.66691242
## 
## ------------------------------------------------------
## * 95 percent confidence interval
## ------------------------------------------------------
\end{verbatim}
\end{kframe}
\end{knitrout}

Considering only term births in Example 1 seems to lead to an underestimation of the difference in proportion of low birthweight babies between smoking and non-smoking mothers. However, we have correlation in the observations that we did not take into account. This is done in the next example.

\subsubsection{Example 9}

The last example illustrates the distributional method applied to  hierarchical structures. For this we keep our birthweight-smoke example, but we control for twins. It has been shown \cite{sauzet2013}, that even with low percentage of twins, better estimates of parameters are obtained when the non-independence of twins was accounted for using a multilevel model. The hierarchical structure consists of the baby at level one and the mother at level 2. We fitted a random intercept model.

\begin{knitrout}\footnotesize
\definecolor{shadecolor}{rgb}{0.969, 0.969, 0.969}\color{fgcolor}\begin{kframe}
\begin{alltt}
\hlkwd{library}\hlstd{(distdichoR)}
\hlkwd{library}\hlstd{(nlme)}
\hlstd{mod_smoke} \hlkwb{<-} \hlkwd{lme}\hlstd{(birthwt} \hlopt{~} \hlstd{smoke2} \hlopt{+} \hlstd{gest,} \hlkwc{data} \hlstd{= bwsmokecompl,} \hlkwc{na.action} \hlstd{= na.omit,}
                 \hlkwc{random} \hlstd{=} \hlopt{~} \hlnum{1} \hlopt{|} \hlstd{momid,} \hlkwc{method} \hlstd{=} \hlstr{'ML'}\hlstd{)}
\hlkwd{summary}\hlstd{(mod_smoke)}
\end{alltt}
\begin{verbatim}
## Linear mixed-effects model fit by maximum likelihood
##  Data: bwsmokecompl 
##        AIC      BIC    logLik
##   29634.28 29668.13 -14811.14
## 
## Random effects:
##  Formula: ~1 | momid
##         (Intercept) Residual
## StdDev:    420.2496 36.76134
## 
## Fixed effects: birthwt ~ smoke2 + gest 
##                  Value Std.Error   DF   t-value p-value
## (Intercept) -2729.6149 189.85796 1728 -14.37714   0e+00
## smoke21      -165.3941  22.62879 1728  -7.30902   0e+00
## smoke22      -128.2181  36.41752 1728  -3.52078   4e-04
## gest          154.6594   4.79921 1728  32.22599   0e+00
##  Correlation: 
##         (Intr) smok21 smok22
## smoke21 -0.046              
## smoke22 -0.054  0.204       
## gest    -0.998  0.007  0.030
## 
## Standardized Within-Group Residuals:
##          Min           Q1          Med           Q3          Max 
## -1.895862639 -0.088679885 -0.005289763  0.092746551  1.891732633 
## 
## Number of Observations: 2083
## Number of Groups: 1732
\end{verbatim}
\begin{alltt}
\hlkwd{regdistdicho}\hlstd{(}\hlkwc{mod} \hlstd{= mod_smoke,} \hlkwc{group_var} \hlstd{=} \hlstr{'smoke2'}\hlstd{,} \hlkwc{cp} \hlstd{=} \hlnum{2500}\hlstd{)}
\end{alltt}
\begin{verbatim}
## [[1]]
## ======================================================
## ===              Distributional method             ===
## ======================================================
## Distributional estimates for the comparison of proportions below the cut-point 2500 
## Standard error computed under the hypothesis that the ratio of variances is equal to 1 
## 
##  Group  Obs     Mean  Std.Dev Dist.prop.
##      1  631 3125.941 421.8544 0.06893314
##      0 1287 3291.335 421.8544 0.03033805
## 
## ------------------------------------------------------
##        Stat   Estimate     Std.Err  CI.lower   CI.upper
##  Diff. prop 0.03859509 0.005618465 0.0275831 0.04960708
##  Risk ratio 2.27216776 0.227221909 1.8704463 2.76016817
##  Odds ratio 2.36635489 0.250339252 1.9265087 2.90662344
## 
## ------------------------------------------------------
## * 95 percent confidence interval
## ------------------------------------------------------
## [[2]]
## ======================================================
## ===              Distributional method             ===
## ======================================================
## Distributional estimates for the comparison of proportions below the cut-point 2500 
## Standard error computed under the hypothesis that the ratio of variances is equal to 1 
## 
##  Group  Obs     Mean  Std.Dev Dist.prop.
##      2  188 3163.117 421.8544 0.05798581
##      0 1287 3291.335 421.8544 0.03033805
## 
## ------------------------------------------------------
##        Stat   Estimate     Std.Err   CI.lower   CI.upper
##  Diff. prop 0.02764776 0.008672217 0.01065053 0.04464499
##  Risk ratio 1.91132289 0.309583146 1.39975269 2.60985759
##  Odds ratio 1.96741949 0.337530789 1.41554784 2.73444624
## 
## ------------------------------------------------------
## * 95 percent confidence interval
## ------------------------------------------------------
\end{verbatim}
\end{kframe}
\end{knitrout}

Not accounting for correlated twins leads to an underestimation of the estimates for the comparison of proportions. However, the standard errors for these estimates are not affected.

\section{Conclusion}

The functions available in the package \code{distdichoR} make the distributional method for the dichotomisation of continuous outcomes easily accessible either for simple comparison following a t-test or to obtain adjusted comparisons for a range of distributions.
\vskip 180pt
{\bf Funding}: \\
We acknowledge the financial contribution granted  for this work from the Research Centre for Mathematical Modelling, Bielefeld University. \\

{\bf Affiliation}:\\
Odile Sauzet\\ Statistical Consulting Centre, Centre for Statistics \\
\& \\ Department of Epidemiology \& International Public Health,\\ Bielefeld School of Public Health (BiSPH),\\  Bielefeld University, Bielefeld, Germany\\
odile.sauzet@uni-bielefeld.de



\bibliography{ref}
\bibliographystyle{vancouver}

\end{document}